\documentclass{emulateapj}
\usepackage{apjfonts}
\usepackage{graphicx}

\def\24m{24 $\mu$m}

\def\kpc{$h^{-1}$kpc }
\def\kms{${\rm km~s^{-1}}$ }

\def\dis{$r_{p}$ }
\def\vel{$\vartriangle$$v$ }

\def\mbe{$M_{B}^{e}$ }
\def\abm{$M_{B} $}
\def\Nc{$N_{c}$ }
\def\Ncb{$N_{c}^{b}$ }
\def\Ncr{$N_{c}^{r}$ }
\def\Ncm{$N_{c}^{m}$ }
\def\Nmg{$N_{mg}$ }
\def\Nmgw{$N_{mg}^{wet}$ }
\def\Nmgd{$N_{mg}^{dry}$ }
\def\Nmgm{$N_{mg}^{mix}$ }

\shorttitle{WET, DRY, AND MIXED GALAXY MERGERS IN DEEP2}

\shortauthors{Lin et al.}

\begin{document}

\title{The Redshift Evolution of Wet, Dry, and Mixed Galaxy Mergers from Close Galaxy Pairs in the DEEP2 Galaxy Redshift Survey}

\author{Lihwai Lin \altaffilmark{1,2}, David R. Patton \altaffilmark{3}, David C. Koo \altaffilmark{2}, Kevin Casteels
\altaffilmark{3}, Christopher J. Conselice \altaffilmark{4}, S. M. Faber \altaffilmark{2}, Jennifer Lotz
\altaffilmark{5,6}, Christopher N. A. Willmer \altaffilmark{7}, B. C. Hsieh
\altaffilmark{1}, Tzihong Chiueh \altaffilmark{8}, Jeffrey A. Newman \altaffilmark{9}, Gregory S. Novak
\altaffilmark{2}, Benjamin J. Weiner \altaffilmark{7}, Michael C. Cooper \altaffilmark{7,10}}

\altaffiltext{1}{Institute of Astronomy \& Astrophysics,
Academia Sinica, Taipei 106, Taiwan; Email: lihwailin@asiaa.sinica.edu.tw} \altaffiltext{2}{UCO/Lick Observatory, Department of Astronomy and Astrophysics, University of California, Santa
Cruz, CA 95064} \altaffiltext{3}{Department of Physics and Astronomy, Trent University, 1600
West Bank Drive, Peterborough, ON K9J 7B8 Canada} \altaffiltext{4}{School of Physics and Astronomy, University of
Nottingham, Nottingham, NG72RDUK} \altaffiltext{5}{National Optical Astronomy Observatory, 950 N. Cherry Ave.,
Tucson, AZ 85719} \altaffiltext{6}{Leo Goldberg Fellow} \altaffiltext{7}{Steward Observatory, University of Arizona,
933 N.\ Cherry Avenue, Tucson, AZ 85721 USA} \altaffiltext{8}{Department of Physics, National Taiwan University,
Taipei, Taiwan} \altaffiltext{9}{Physics and Astronomy Dept., University of Pittsburgh, Pittsburgh, PA, 15620}
\altaffiltext{10}{Spitzer Fellow}

\begin{abstract}
\hspace{3mm} We study the redshift evolution of galaxy pair fractions and merger rates for different types of
galaxies using kinematic pairs selected from the DEEP2 Redshift Survey, combined with other surveys at lower
redshifts. By parameterizing the evolution of the pair fraction as $(1+z)^{m}$, we find that the companion rate
increases mildly with redshift with $m = 0.41\pm0.20$ for all galaxies with $-21 <$ \mbe $< -19$. Blue
galaxies show slightly faster evolution in the blue companion rate with $m = 1.27\pm0.35$, while red galaxies have
had fewer red companions in the past as evidenced by the negative slope $m = -0.92\pm0.59$. The different trends of
pair fraction evolution are consistent with the predictions from the observed evolution of galaxy number densities
and the two-point correlation function for both the blue cloud and red sequence. For the chosen luminosity range, we find that at low redshift the pair fraction within the red sequence exceeds that of the blue cloud, indicating a higher merger probability among red galaxies compared to that among the blue galaxies. With further assumptions on the
merger timescale and the fraction of pairs that will merge, the galaxy major merger rates for $0.1 < z <
1.2$ are estimated to be
$\sim~10^{-3}$ $h^{3}$Mpc$^{-3}$Gyr$^{-1}$ with a factor of 2 uncertainty. At $z \sim 1.1$, 68\%
of mergers are wet, 8\% of mergers are
dry, and 24\% of mergers are mixed, compared to 31\% wet mergers, 25\% dry mergers, and 44\% mixed mergers at $z
\sim 0.1$. Wet mergers dominate merging events at $z = 0.2 - 1.2$, but the relative importance of dry and mixed
mergers increases over time.
The growth of dry merger rates with decreasing redshift is mainly due to the increase in the
co-moving number density of red galaxies over time. About 22\% to 54\% of present-day $L^{*}$ galaxies have
experienced major mergers since $z \sim 1.2$, depending on the definition of major mergers. Moreover, 24\% of the
red galaxies at the present epoch have had dry mergers with luminosity ratios between 1:4 and 4:1 since $z \sim 1$. Our results also suggest that all three
types of mergers play an important role in the growth of the red sequence, assuming that a significant fraction of
wet/mixed mergers will also end up as red galaxies. However, the three types of mergers lead to red galaxies in
different stellar mass regimes: the wet mergers and/or mixed mergers may be partially responsible for producing red galaxies
with intermediate masses while a significant portion of massive red galaxies are assembled through dry mergers at
later times.

\end{abstract}


\section{INTRODUCTION}
According to the $\Lambda$-dominated Cold Dark Matter ($\Lambda$CDM) model, major mergers of galaxies are an
important process in the formation of present-day massive galaxies. The merger rate of dark matter halos and
galaxies as well as its evolution have now been widely studied with $N-$body simulations and semi-analytical models
\citep{lac93,gov99,got01,kho01,mal06,ber06,fak07,guo08,mat08}.
Measuring the frequency of galaxy close pairs and galaxy merger rates thus provides powerful constraints on
theories of galaxy formation and evolution. The galaxy merger fraction is often parameterized by a power law of
the form $(1+z)^{\rm m}$.  Observational studies of galaxy merger rates using both close pairs and morphological
approaches within the last decade have found a diverse range of $m$ values from $m \sim 0$ to $\sim 4$
\citep{zep89,bur94,car94,yee95,woo95,neu97,pat97,car00,le00,pat02,con03,bun04,lin04,cas05,con06,bel06b,lot08,kam07,kar07}.
This discrepancy may arise from the different sample selections across different redshift ranges, as
well as from different procedures used to correct for the sample incompleteness. The relatively mild evolution
of observed
galaxy mergers found in the literature \citep{car00,bun04,lin04,lot08} seems to be in contradiction to the rapid increase of halo merger rates with redshift predicted in $N-$body numerical simulations where $m \sim 3$ \citep{gov99,got01}. Nevertheless, such comparison may not be adequate since the latter were focused on the merger histories of distinct halos which host one or multiple galaxies. On the other hand, the mergers of subhalos in $N-$body simulations offer a better analogy to the observed galaxy mergers. In a recent study
using $N-$body simulations, \citet{ber06} find that the companion rate of subhalos increases mildly with
redshift out to $z \sim 1$, consistent with the data presented in \citet{lin04}.

Despite the successful agreement between pair counts of observed galaxies and subhalos in simulations, it is not yet
clear whether low or mild evolution of the pair fraction and merger rate still holds for different types of
galaxies. The intrinsic color distribution of galaxies has been shown to be bi-modal since $z \sim 1$
\citep{bel04,fab07}. It is thus expected that the effects of interactions and mergers between various types of
galaxies on their final products can be different. For example, 'wet mergers' (mergers between two gas-rich
galaxies) can trigger additional star formation \citep{bar00,lam03,nik04,woo06,lin07,bri07,bar07}, cause quasar activity
\citep{hop06} and transform disk galaxies into ellipticals \citep{too72}. On the other hand, the so-called 'dry
mergers' (mergers between two gas-poor galaxies) may not involve dramatic changes in the star formation rate, but
can play an important role in the stellar mass growth of massive red galaxies at the current epoch
\citep{tra05,van05,bel06a,fab07,mci07,kho03,kho05,naa06,cat08}.
In addition, the relative fraction of mixed pairs
versus separation might yield clues on the effectiveness of a red galaxy to shut down the star
formation even of other galaxies in its neighborhood. Quantifying the merger rates of galaxies between different
types is therefore an important step towards understanding how present day massive galaxies are built up.

While recently there have been several efforts attempting to estimate the merger rates of certain categories, they were
mainly focused on red galaxies
\citep{van05,bel06a,mas06,lot08,mas07}. Yet no direct observational measurement of the relative abundances of wet, dry, and mixed mergers has been provided. despite that they have been explored in recent theoretical studies \citep{kho03,cio07}.
In this work, we investigate the evolution of the pair fractions for different types of galaxies and obtain the
relative fraction of major merger rates among various types of mergers for the first time. We classify close galaxy pairs into three
different categories (blue-blue pairs; red-red pairs; mixed pairs) based on galaxy colors. As a first
approximation, blue galaxies are gas-rich while red galaxies are gas-poor. We therefore
calculate the merger rates of wet mergers, dry mergers, and mixed mergers using the number statistics from blue-blue
pairs, red-red pairs and mixed pairs. There is however possible contamination of red gaseous and blue gas-poor
galaxies in our analysis. Recent studies of galaxy morphologies have suggested that about 20\% of red galaxies
appear to be either edge-on disks or dusty galaxies and hence are likely to be gas-rich \citep{wei05}. On the
other hand, there also exist blue spheroidals that could be gas-poor, although these are relatively rare objects
\citep{cas07}. Since both cases of contamination discussed above affect only a minority of the red sequence and
blue clouds respectively, classifying different types of mergers based on their colors should be a good approximation. The pair sample is constructed based on their rest-frame $B$-band luminosity $L_{B}$ as often used in the literature. While the stellar mass range of red and blue galaxies selected with fixed $L_{B}$ could be different, $L_{B}$ is shown to be a good tracer of dynamical mass for a wide range of Hubble types \citep{kan08}. Therefore, in this work we select major-merger candidates based on the ratio of $L_{B}$ regardless of the color difference.

The galaxy sample at $0.45 < z < 1.2$ is taken from the DEEP2 Redshift Survey \citep{dav03,dav07} and Team Keck
Redshift Survey in GOODS-N \citep{wir04}. We also supplement our low redshift sample (z $<$ 0.45) using the SSRS2
survey \citep{dac98}, Millennium Galaxy Catalog \citep[hereafter MGC,][]{lis03,dri05,all06} and CNOC2 Redshift
Survey \citep{yee00}. The combined samples yield the largest number of kinematic pairs out to $z \sim 1.2$ to date
and enable study, for the first time, of galaxy merger rates for wet mergers, dry mergers and mixed mergers as a function of redshift.

In \S 2, we describe the selection of close pairs. In \S3, we present our results on the pair fractions for blue and
red galaxies, as well as the derived merger rates for different merger categories. A discussion is given in
\S4, followed by our conclusions in \S 5. Throughout this paper we adopt the following cosmology: H$_0$ = 100$h$~\kms Mpc$^{-1}$, $\Omega_m =
0.3$ and $\Omega_{\Lambda } = 0.7$. The Hubble constant $h$ = 0.7 is adopted when calculating rest-frame
magnitudes. Unless indicated otherwise, magnitudes are given in the AB system.

\section{DATA, SAMPLE SELECTIONS, AND METHODS}
Close pairs are potential progenitors of merging galaxies and hence present an opportunity to study the different
types of mergers before coalescence takes place. Thanks to the high spectral resolution of DEEP2 ($\sim$ 30 \kms)
and TKRS ($\sim$ 60 \kms), we are able to select kinematic pairs at $0.45 < z < 1.2$, which require accurate
spectroscopic redshifts of both pair components in order to reduce the contamination by interlopers. Three other
redshifts surveys including galaxies at lower redshift ($z < 0.5$) - SSRS2 , MGC, and CNOC2 - are also added to our
sample.
\subsection{K-correction and Sample selection}
The rest-frame $B$-band magnitudes (\abm) and $U - B$ colors for DEEP2 galaxies at $0.45
< z < 0.9$ are derived in a similar way to that in \citet{wil06}. For galaxies with $0.9 < z <
1.2$, the rest-frame $U - B$ color is computed using the observed $R - z_{\mathrm{mega}}$ color, where
$z_{\mathrm{mega}}$ is the $z$-band magnitude obtained from CFHT/Megacam observations for DEEP2 Fields in 2004 and
2005 (Lin et al. 2008, in preparation). The $K$-corrections for the TKRS sample are described in \citet{wei06}.

We started from a sample of galaxies with $-21 <$ \mbe $< -19$, where \mbe
is the evolution-corrected absolute magnitude, defined as \abm + $Qz$. The values of $Q$ are found to be close to 1.3 up to $z \sim 1$ for either blue or red galaxies \citep{fab07}. Throughout this paper, we therefore adopt $Q = 1.3$
to ensure that galaxies within the same range of the luminosity function are being selected. Kinematic close pairs
are then identified such that their projected separations satisfy 10 \kpc $\leq$ \dis
$\leq$ $r_{max}$ (physical length) and rest frame relative velocities \vel less than 500 \kms \citep{pat00,lin04}.

\begin{figure}
\includegraphics[angle=-90,width=9.5cm]{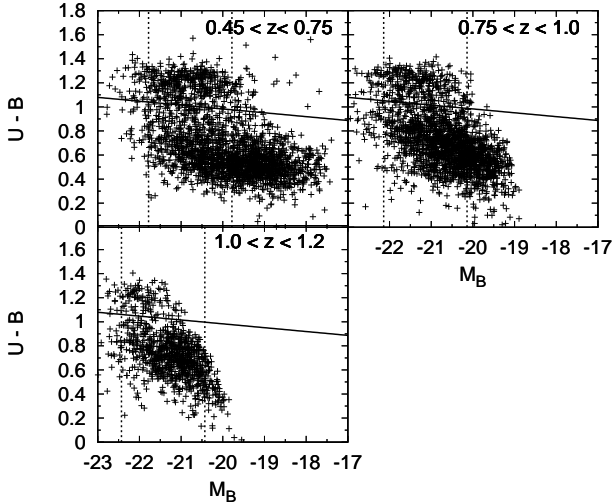}
%\includegraphics[angle=-90,width=9.5cm]{cmd_deep2.eps}
%\epsscale{.7}
%\plotone{f2.eps}
\caption{Restframe $U - B$ vs. absolute $B$-band magnitude for galaxies in one of the DEEP2 fields (EGS).
The solid lines denote the color cut (see Equation \ref{color}) to separate the blue and red galaxy
populations. The vertical dotted lines in each panel indicate the approximate bright and faint limit of
absolute $B$
magnitude ($M_{B}$) at the mean redshift of each redshift range used to pick up pair samples.
\label{figcmd_1}} \end{figure}

Galaxies are further divided into the blue cloud and red sequence using the rest-frame magnitude dependent cut for
DEEP2 and TKRS (in AB magnitudes):
\begin{equation}\label{color}
U - B = -0.032(M_{B} + 21.62) + 1.035.
\end{equation}
Fig. \ref{figcmd_1} shows the rest-frame color-magnitude diagram for one of the DEEP2 fields (EGS) in three redshift bins. The solid lines denote the above color cut to separate the blue and red galaxies. The vertical dotted lines in each panel indicate the approximate bright and faint limit of $M_{B}$ corresponding to $-21 <$ \mbe $< -19$. It can be seen that the red galaxies are not complete in the highest redshift bin ($bottom$ $left$) due to the $R = 24.1$ cut in the DEEP2 sample. We will discuss how to deal with such incompleteness in \S 3.1.

\begin{figure}
\includegraphics[angle=-90,width=9cm]{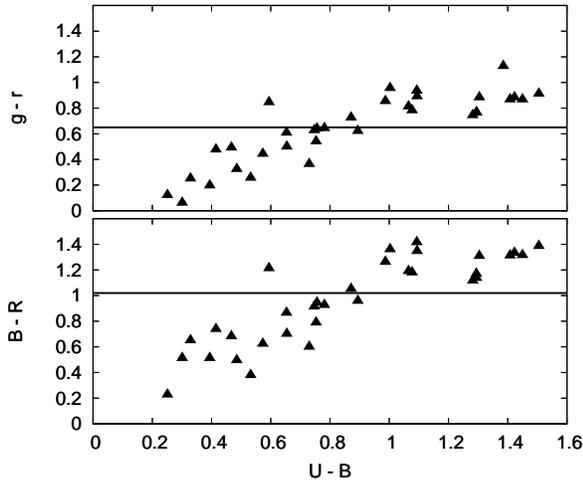}
%\includegraphics[angle=-90,width=9cm]{BR-gr-UB.eps}
%\epsscale{.7}
%\plotone{f2.eps}
\caption{Synthesized $g - r$ vs. $U - B$ color and $B - R$ vs. $U - B$ color from galaxy templates of Kinney
et al. (1996). It can be seen that both the rest-frame $g - r$ and $B - R$ colors correlate well with the
rest-frame $U - B$ color. The solid lines denote the color thresholds adopted in our work to separate the
blue and red
galaxy populations for MGC and CNOC2 samples respectively. \label{figcolor}} \end{figure}

\begin{figure}
\includegraphics[angle=-90,width=9cm]{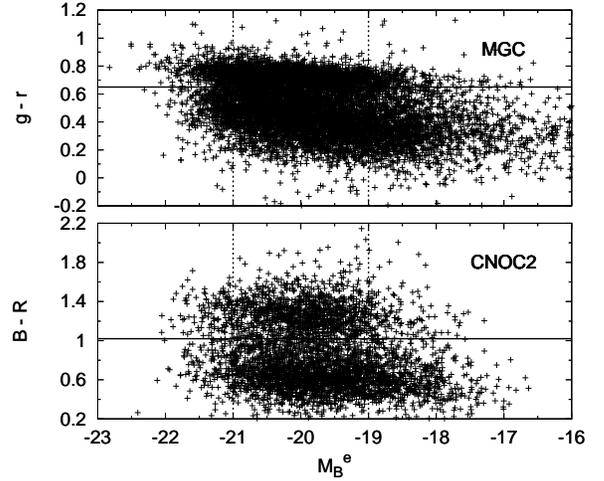}
%\includegraphics[angle=-90,width=9cm]{cmd_MGCCNOC2.eps}
%\epsscale{.7}
%\plotone{f2.eps}
\caption{Upper: rest-frame $g - r$ vs. evolution-corrected absolute $B$-band magnitude ($M_{B}^{e}$)
diagram. Lower panel: rest-frame $B - R$ vs. evolution-corrected absolute $B$-band magnitude ($M_{B}^{e}$).
The solid lines denote the color thresholds adopted in our work to separate the blue and red galaxy
populations. The vertical dotted lines in each panel indicate the bright and faint limit of $M_{B}^{e}$
used to select samples for pair studies. \label{figcmd_2}} \end{figure}

For the low redshift samples, simple rest-frame color cuts $g - r = 0.65$ (in AB) and $B-R = 1.02$
(in AB) are applied to MGC and CNOC2
respectively. In Fig. \ref{figcolor}, we plot the relation between the rest-frame $g - r$ and $U - B$ ($top$) and between the rest-frame $B - R$ and $U - B$ ($bottom$) using the synthesized colors from templates of \citet{kin96}. As shown in Fig. \ref{figcolor}, there is a fairly good correlation between these colors. Therefore the $g - r$ cut for MGC and the $B - R$ cut for CNOC2 can still provide good correspondence of blue and red galaxies at low-redshifts to the DEEP2 sample. Fig. \ref{figcmd_2} shows the color versus the evolution-corrected $B$-band magnitude for MGC and CNOC2 ($top$ $and$ $bottom$, $respectively$), with the two dotted lines corresponding to the $-21 <$ \mbe $< -19$ cut. Blue-blue pairs, red-red pairs, and mixed pairs (hereafter b-b, r-r, and mixed pairs respectively) are
classified according to the color combination of the pairs. In total, we have 218 b-b pairs, 122 r-r pairs, and 166
mixed pairs with 10 \kpc $\leq$ \dis $\leq$ $50$ \kpc and \vel $\leq$ 500 \kms from combined samples.

\subsection{The Spectroscopic Selection Function and Weights}
To measure the incompleteness of the DEEP2 survey and hence the selection function, we compared the sample
with successful redshifts to all objects in  the photometric catalog that satisfy  the limiting magnitude and any
photometric redshift cut. The selection function is expected to depend on an unknown and complex interplay among
observables and intrinsic properties of objects. With data too limited  to undertake multi-dimensional
investigations of the selection function, we make the simplifying assumption that the selection function is
separable in the different observed variables \citep{yee96}. By assuming that fluxes are the only observables correlated
to the other observables of galaxies, we restrict the definition of the  selection function to be
\begin{equation}\label{weight}
S = S_{m}\;\overline{S_{c}}\;\overline{S_{SB}}\;\overline{S_{xy}} =
S_{m}(R)\frac{S_{c}(B-R,R-I,R)}{S_{m}(R)}\frac{S_{SB}(\mu_{R},R)}{S_{m}(R)}\frac{S_{xy}}{S_{m}(R)},
\end{equation}
where $S_{m}$ is the magnitude selection function, $S_{c}$ is the apparent color selection function, $S_{SB}$ is the
surface brightness selection function and $S_{xy}$ represents the geometric (local density) selection
function. $\overline{S_{c}}$, $\overline{S_{SB}}$, and $\overline{S_{xy}}$ are all normalized to the magnitude selection function, $S_{m}$. The
spectroscopic weight $w$ for each galaxy is  thus $1/S$, which is derived from its apparent $R$ mag, $B-R$ and $R-I$  colors,
$R$ band surface brightness, and local galaxy density.

\begin{figure}
\includegraphics[angle=-90,width=8.5cm]{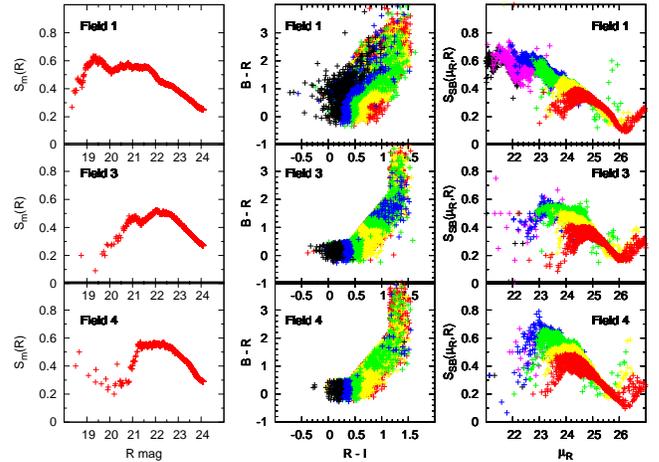}
%\includegraphics[angle=-90,width=8.5cm]{SmScSB.eps}
%\epsscale{.7}
%\plotone{f2.eps}
\caption{(1) Left: the apparent $R$-band magnitude selection $S_{m}(R)$ as a function of the
apparent
$R$-band
magnitude. The peak completeness is about 55\% $\sim$ 60\%. (2) Middle: the
apparent color selection function $S_{c}(B-R,R-I,R)$ as a function the apparent color $B
- R$ and $R - I$ for DEEP2. The colors correspond to various ranges of the
selection function (red is for $S_{c} > 0.5$; yellow is for $0.4 < S_{c} <
0.5$; green is for $0.3 < S_{c} < 0.4$; blue is for $0.2 < S_{c} < 0.3$; black
is for $S_{c} < 0.2$). (3) Right: the
apparent $R$-band surface brightness selection function $S_{SB}(\mu_{R}, R)$ as a function
the apparent $R$-band surface brightness for DEEP2. The colors correspond to the apparent
$R$-band magnitude (red is for $23.5 < R < 24.1$; yellow is for $23 < S_{c} <
23.5$; green is for $22 < R < 23$; blue is for $21 < R < 22$; magenta
is for $20 < R < 21$; black is for $R < 20$).\label{figsf_1}}
\end{figure}

\begin{figure}
\includegraphics[angle=-90,width=9cm]{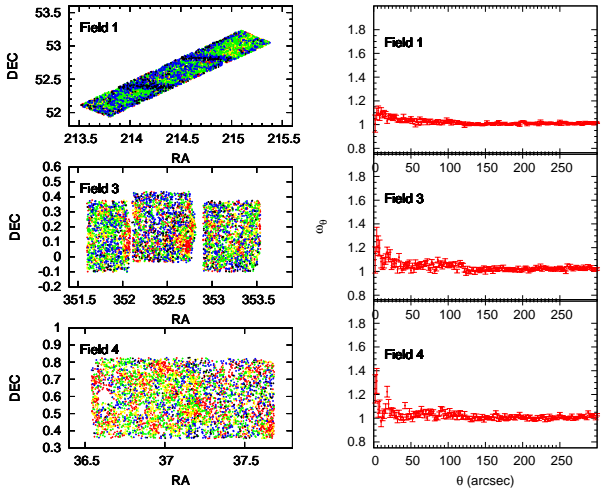}
%\includegraphics[angle=-90,width=9cm]{SxySang.eps}
%\epsscale{.7}
%\plotone{f2.eps}
\caption{Left: The spatial distribution of the geometric selection function for DEEP2 Fields 1, 3, and 4. The colors correspond to various ranges of the
selection function (red is for $S_{c} > 0.5$; yellow is for $0.4 < S_{c} <
0.5$; green is for $0.3 < S_{c} < 0.4$; blue is for $0.2 < S_{c} < 0.3$; black
is for $S_{c} < 0.2$). Right: The angular weights ($\omega_{\theta}$) as a function of angular separation ($\theta$) of pairs. \label{figsf_2}}
\end{figure}

The magnitude selection function $S_{m}(R)$ (the left panel of
Fig. \ref{figsf_1}) of each galaxy is computed as the ratio of
the number of galaxies with good redshift qualities to the total
number of galaxies in the target catalog in both cases considering a
magnitude bin of $\pm$0.25 mag centered on the magnitude of the
galaxy. The color selection function
$S_{c}$($B-R,R-I,R$) (the middle panel of Fig. \ref{figsf_1}) is computed by counting galaxies within $\pm$ 0.25 $R$ magnitude over a
$B-R$ and $R-I$ color range of $\pm$ 0.25 mag. Similarly, the surface brightness selection function (the right panel of Fig. \ref{figsf_1}) is performed within $\pm$0.25 mag in $\mu_{R}$ and $\pm$ 0.25 mag in $R$. The geometric selection function
$S_{xy}$($xy,R$) is similar to magnitude selection function but has a
localized effect. We take the ratio between the number of
galaxies with good quality redshifts and the total number in the targeted catalog in an area of radius 120" within a $\pm$ 0.25 $R$- magnitude
range. The left panel of Fig. \ref{figsf_2} shows the distribution of $S_{xy}$. Finally we use Equation \ref{weight} to compute the total selection function $S$, which leads to the spectroscopic weight for each galaxy in DEEP2 as $w = 1/S$.

Besides the selection function for each individual galaxy, we also investigate the selection dependence on pair separation using analogous procedures adopted by \citet{pat02}. In
principle, the target selection is unlikely to place slits on close
pairs simultaneously since the slit orientations constrained to be
less than $\pm 30$ degs from the slit mask orientation. In addition, we are not able to put slits on objects that are very close to
each other because their separate spectra will overlap. The suppression of close pairs, however, is not a severe problem in
DEEP2 because each field has been observed with two masks. To quantify this effect, we measure the angular separation of all pairs in the redshift
catalog (z-z pairs) and in the target catalog (p-p pairs) respectively and then count the number of pairs ($N_{zz}$
and $N_{pp}$) within each angular separation bin. While counting the pairs in the redshift catalog, each component
of a pair is weighted by a geometric selection function $S_{xy}$($xy$) to exclude the effect due to the
variance in the local
sampling rate. The angular selection function $S_{\theta}$ is computed as the ratio between the weighted $N_{zz}$ and $N_{pp}$. The angular weight, $w_{\theta}$, for each galaxy is hence $1/S_{\theta}$ (see the right panel of Fig. \ref{figsf_2}).

We repeat the above analysis for the TKRS sample except that the ACS $B - V$ and $V - i$ colors are used when calculating the color selection function. The selection function and weights for SSRS2, MGC, and CNOC2 samples are computed in the same manner as described in \citet{pat00,pat02}.
\section{RESULTS}

\begin{figure*}
\includegraphics[angle=-90,width=17cm]{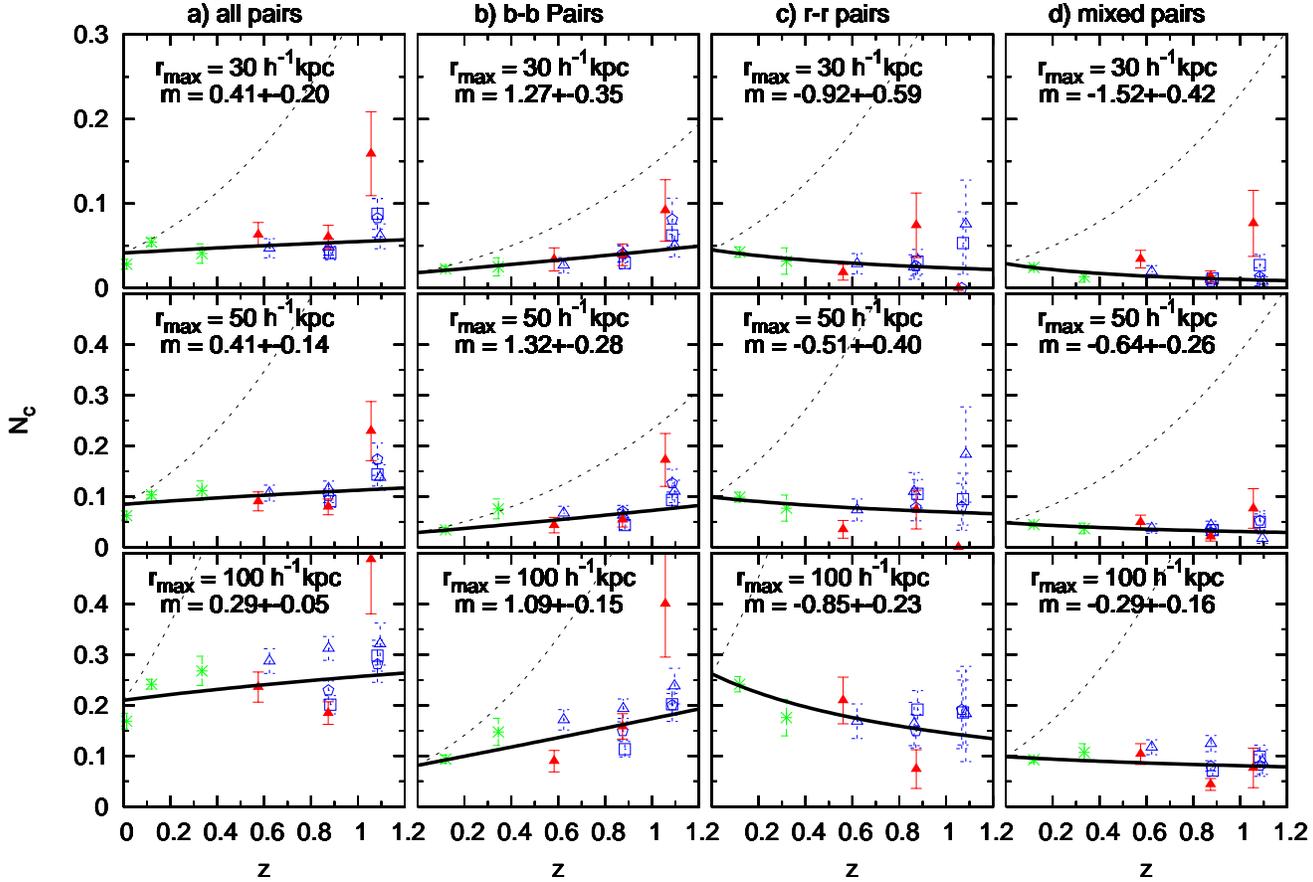}
%\includegraphics[angle=-90,width=17cm]{Nc-z_alltype.eps}
%\epsscale{.7}
%\plotone{f1.eps}
\caption{The pair fraction as a function of redshift for different types of close
pairs using $r_{max}= $30 $h^{-1}$ kpc (top panel), 50 $h^{-1}$ kpc (middle panel), and  100 $h^{-1}$ kpc (bottom
panel). From left to right: all pairs, b-b pairs, r-r- pairs, and mixed pairs. Colors represent data points from
different surveys: green for the SSRS2 ($\overline{z}$ $\sim$ 0.01), MGC ($\overline{z}$ $\sim$ 0.12), and CNOC2
($\overline{z}$ $\sim$ 0.34); blue for the DEEP2 Fields 1 (blue triangles), 3 (squares), and 4 (pentagons); red is
for TKRS. The points lying on the X-axis represent the fields where no pairs have been found; they are not included
when calculating the fits. The best fits are shown as solid lines while the dotted lines represent the $m = 3$
curves. Different types of pairs evolve differently as a function of redshift denoted by the value of evolution
power $m$ shown on each plot. The error bars shown in the plot and
used for fitting are calculated by bootstrapping. \label{figNc}}
\end{figure*}

\begin{figure*}
\includegraphics[angle=-90,width=17cm]{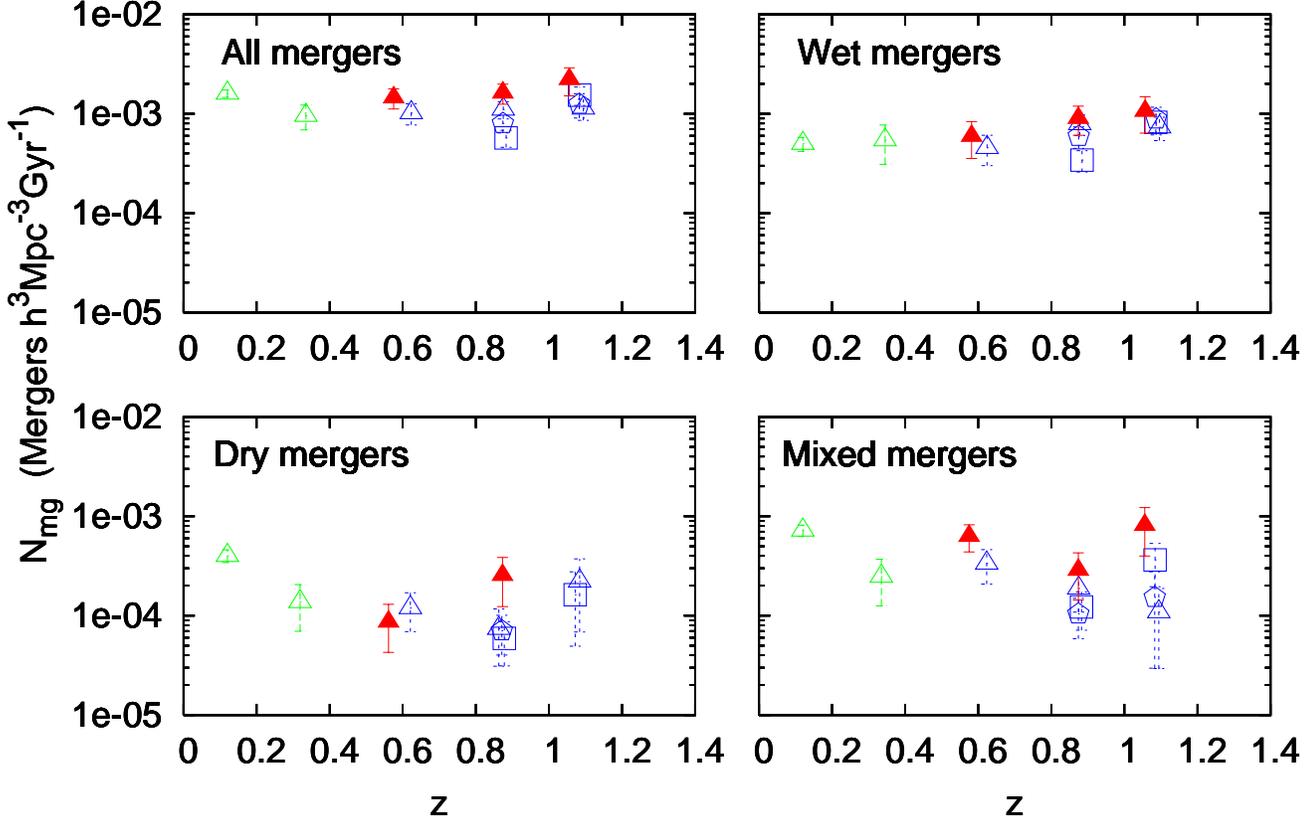}
%\includegraphics[angle=-90,width=17cm]{Nm-z_alltype.eps}
%\epsscale{.7}
%\plotone{f2.eps}
\caption{Comoving volume major merger rate as a function of redshift for the
various types of mergers as indicated in the plots. Different symbols represent data from different survey fields as
described in Fig.~\ref{figNc}. The errors shown here
represent the uncertainty coming from the pair counts
in our samples, and do not include the uncertainties of $T_{mg}$
and $C_{mg}$. \label{figNm}}
\end{figure*}

\begin{figure}
\includegraphics[angle=-90,width=8cm]{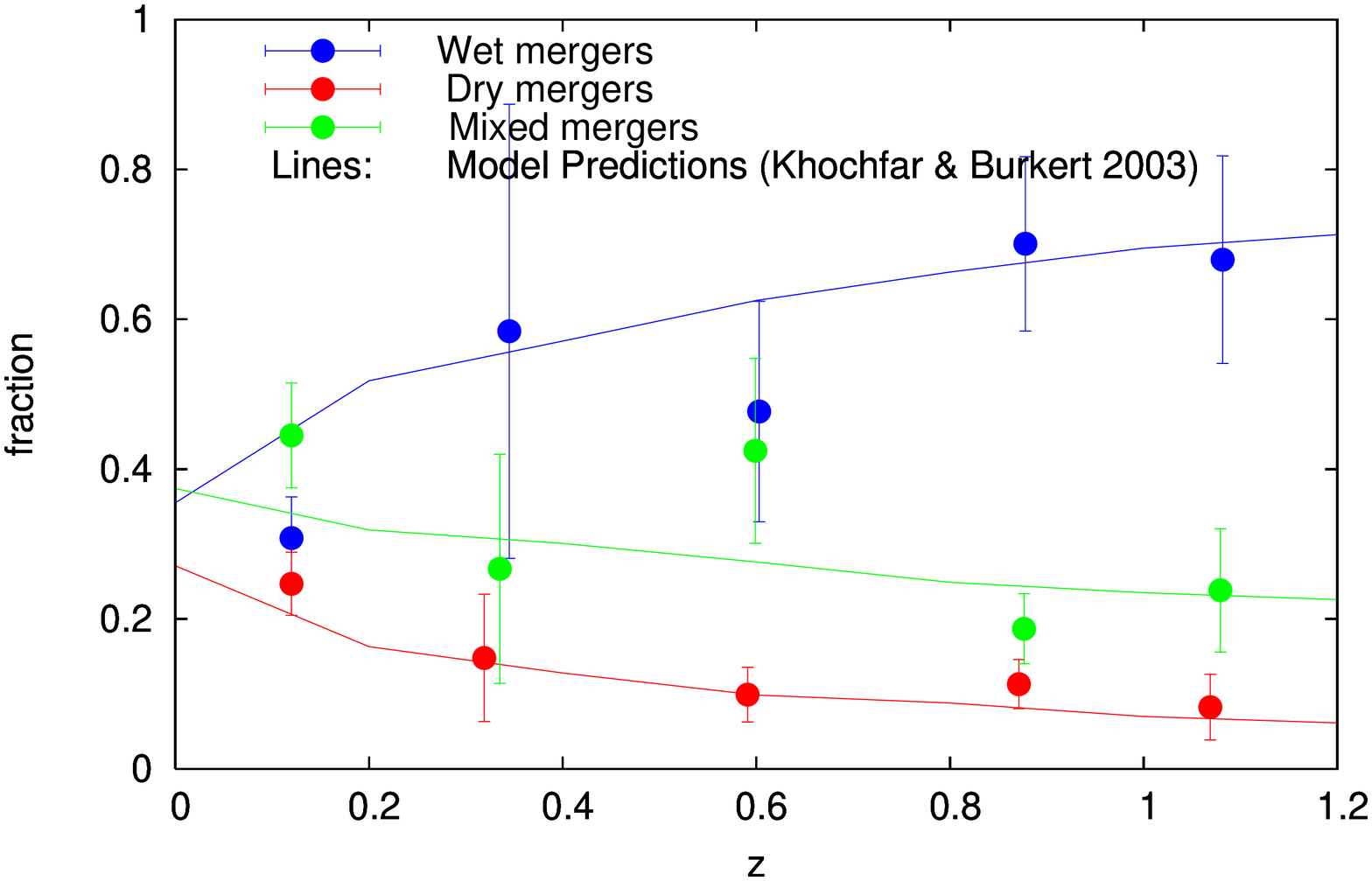}
%\includegraphics[angle=-90,width=8cm]{Nmratio-z_alltype_1p_1e13_ave.eps}
%\epsscale{.7}
%\plotone{f3.eps}
\caption{Fraction of major mergers for wet (blue symbols), dry (red symbols), and
mixed mergers (green symbols) as a function of redshift. The data points represent results from the DEEP2, TKRS, CNOC2
and MGC surveys. The three color lines show the semi-analytical predictions of Sp-Sp, E-E, and E-Sp mergers by
\citet{kho03} but for a field-like environment, corresponding to a dark matter halo of mass $M_{0} \sim 10^{13}
M_{\odot}$. The data and the model prediction are in good agreement, both showing an increasing fraction of dry
(E-E) and mixed (E-Sp) mergers with decreasing redshift. \label{figNmratio}}
\end{figure}

\subsection{The Pair Fraction for the Blue Cloud, the Red Sequence and Full Galaxy Sample}
We first compute the pair fraction \Nc, defined as the average number of companions per galaxy:
\begin{equation}\label{Nc}
N_{c}=\frac{\sum^{N_{tot}}_{i=1}\sum_{j}w_{j}w(\theta)_{ij}}{N_{tot}},
\end{equation}
where $N_{tot}$ is the total number of galaxies within the chosen absolute magnitude range, $w_{j}$ is the
spectroscopic weight for the $j$th companion belonging to the $i$th galaxy, and $w(\theta)_{ij}$ is the angular
selection weight for each pair as described in \S 2.2. While the blue galaxy sample is volume-limited for the adopted
magnitude range $-21 <$ \mbe $< -19$, because of the  $R = 24.1$ cut in the DEEP2 sample, lower luminosity red
galaxies beyond $z \sim 1$ are not contained in the sample. To estimate the missing fraction of galaxies, we extrapolated the derived red luminosity function of DEEP2 \citep{wil06} and calculated the expected total number of galaxies between
 -21 $<$ \abm + $Qz$ $<$ -19 at $z \sim 1.1$. These predicted numbers are then compared to the
predicted number of galaxies between -21 $<$ \abm + $Qz$ $<$ $M_{limit}$(24.1), where $M_{limit}$(24.1) represents
the faint limit imposed by the apparent R=24.1 cut used by DEEP2. The latter comprise $\sim$ 42\% from which we
estimate that about 58\% of the galaxies could be missing in the sample. Therefore we apply a correction factor of
2.4 for each red companion at $z > 1$ in addition to the usual spectroscopic and angular separation corrections.

Fig.~\ref{figNc} (see also Table~\ref{tbl-1}) shows \Nc versus redshift with $r_{max}= $30 $h^{-1}$ kpc, 50
$h^{-1}$
kpc, and 100 $h^{-1}$ kpc ($from$ $top$ $to$ $bottom$, $respectively$) from four types of measurement: a) \Nc from all pairs regardless of colors;
b) the average number of blue companions per blue galaxy \Ncb; c) the average number of red companions per red
galaxy  \Ncr; d) the average number of companions of galaxies with opposite colors to that of the primary galaxies
\Ncm. Types b) and c) are equivalent to the pair fraction within the blue cloud and red sequence respectively. If the pair
fraction is fitted by $N_{c}(0)(1+z)^{m}$ with both \Nc(0) and $m$ as free parameters, we find that $m$
varies among
different color samples (Table~\ref{tab2}). When considering all colors together, we find a minor amount of
evolution with a power-law index of $m = 0.41\pm0.20$ for the case of $r_{max}=$ 30 \kpc, and $m = 0.41\pm0.14$ for
the case of $r_{max}=$ 50 \kpc. These results are consistent with the value $0.51\pm0.28$ given in \citet{lin04},
which used a sample 7 times smaller. Blue galaxies, however, have stronger evolution $m = 1.27\pm0.35$, meaning that
the probability of blue galaxies having a blue companion is higher at higher redshift. Red galaxies, on the other
hand, have higher chance of being found in r-r pairs at lower redshifts than at high redshifts, as indicated by the
negative power index ($m = -0.92\pm0.59$). Finally, \Ncm, which measures the mixed pair fraction, is also found to
decrease with increasing redshifts ($m = -1.52\pm0.42$). Except for the mixed pairs, there is very weak dependence of $m$ on the chosen $r_{max}$, suggesting that our derived $m$ is not strongly affected by the incomplete sampling rate at small scales. On the other hand, the significant change of $m$ for mixed pairs when varying $r_{max}$ may indicate a change in the environment that hosts mixed pairs over time.

The evolutionary trends of blue and red galaxies can be understood as follows: the pair fraction is proportional to
the galaxy number density times the integral at small scales of the real space 2-point correlation function. The
galaxy correlation function is normally approximated by a power law $\xi=(r/r_{0})^{-\gamma}$ at distances ranging
from 0.1 to several Mpc. Under the assumption that the clustering strength at small scales follows the same power
law, the pair fraction inside a physical radius $R$ can be related to the correlation function as
\begin{equation}\label{f_pair}
f_{pair} \sim n_{g}\int_{0}^{R/a}\xi 4\pi r^{2}dr\propto
n_{g}\frac{\gamma_{0}^\gamma}{3-\gamma}R^{3-\gamma}(1+z)^{3-\gamma},
\end{equation}
where $a$ is the expansion factor, $R$ is the maximum separation of close pairs in physical length, and $n_{g} $ is
the comoving galaxy number density. The study of galaxy clustering of DEEP2 galaxies suggests that there has been
little evolution in $\gamma$ for either blue or red galaxies since $z \sim 1$, and $r_{0}$ increases slightly, by
$10\%$ and $15\%$ for blue and red galaxies respectively, from $z\sim 1$ to $z\sim0$ \citep{coi08}. Adopting $\gamma
= 1.64$ (2.06) for blue (red) galaxies at $z = 1$ \citep{coi08}, and accounting for the increase in
number density of blue (red) galaxies by a factor of 1 (2) since $z = 1$ \citep{fab07}, we obtain $m = 1.14$ (-0.48)
for blue (red) galaxies by comparing the pair fraction at $z = 1$ and $z = 0$ using Equation~\ref{f_pair}. These
values of $m$ are fully consistent with what we have found for kinematic pairs of both blue and red galaxies,
indicating that the pair fraction evolution is a natural consequence of evolution of galaxy number density and
galaxy clustering. While consistency with the overall number density and correlation function studies is
encouraging, the close pairs we use here directly probe what is really happening on small scales (ie.
rather than an inward extrapolation of the correlation function).

\subsection{The Major Merger Rates of Wet Mergers, Dry Mergers, and Mixed Mergers}
Here we define the galaxy major merger rates as the number of merger events involving at least one galaxy within
$-21 < $\mbe $< -19$ to merge with another galaxy with luminosity ratio between 4:1 and 1:4 per unit volume per
gigayear. This quantity can be derived from the pair fraction together with the known galaxy number density and the
assumption about the timescale of being pairs before final mergers. It is worth noting, however, the pair fraction
in \S 3.1 is computed using pairs drawn from within a luminosity range of 2 mag. Some true companions may
fall outside the absolute magnitude range of our sample, while some selected companions have luminosity ratios
outside the range of 4:1 to 1:4. To account for both of these effects, we use the following equation to convert the
pair fraction for the case of $r_{max}= $30 \kpc into merger rates \citep{lin04}:
\begin{equation}\label{mg_G}
N_{mg}=(0.5+G)\times n_{g}(z)C_{mg}N_{c}(z)T^{-1}_{mg},
\end{equation}
where $T_{mg}$ is the timescale for close pairs to merge, $C_{mg}$ denotes the fraction of galaxies in close pairs
that will merge within $T_{mg}$, $n_{g}(z)$ is the comoving number density of galaxies, and $G$ is the correction
factor that accounts for the selection effect of companions due to the restricted luminosity range. The factor of
0.5 converts the number of merging galaxies into the number of merger events (i.e., on average, two close companions
correspond to one galaxy pair and hence one merger). The smallest separation pairs with $r_{max}$ = 30 \kpc are
the best tracers of future mergers, and hence our calculations below of merger rates are based on the pair
statistics from pairs with $r_{max}$ = 30 \kpc. We adopt a crude value of $T_{mg}$ = 0.5 Gyr, as suggested by
major merger simulations \citep{con06,lot08b} and C = 0.6, by estimating the fraction
of pairs that are closer than 30 \kpc in real 3-D space among those selected by 10 \kpc $\leq$ \dis $\leq$
$r_{max}$ \kpc and \vel $<$ 500 \kms. It is worth noting that the uncertainty of $T_{mg}$ is at least a factor of 2. Here we make a simple assumption that $T_{mg}$ is the same for all types of mergers. The motivation behind this is that the $B$-band light is a good tracer of dynamical mass \citep{kan08} and hence the merger timescales should be approximately similar for red-red, blue-blue, and mixed pairs when selected with a fixed $M_{B}$ range at a given redshift. The correction factor $G$ for wet and dry mergers is defined as
\begin{equation}\label{G}
1 + G =
\frac{\int_{M_{B}^{min}(z)}^{M_{B}^{max}(z)}n(M,z)dM\int_{M-1.5}^{M+1.5}n(M',z)dM'}{(\int_{M_{B}^{min}(z)}^{M_{B}^{max}(z)}n(M,z)dM)^{2}},
\end{equation}
where $M_{B}^{min}(z)$ = -21 - $Qz$, and $M_{B}^{max}(z)$ = -19 - $Qz$ in our case. Here $n(M,z)$ is the galaxy number
density for galaxies with magnitude $M$ at redshift $z$. The numerator in Equation \ref{G} gives the integrated
number density of the secondary sample used to search for companions with a luminosity ratio between 4:1 and 1:4
relative to the primary galaxies, weighted by the number density of primary galaxies within $-21 <$ \mbe $< -19$
\footnote{The primary galaxy can be either the bright one or the less luminous one in pairs}. The denominator
gives the integrated number density of companions with $-21 <$ \mbe $< -19$ weighted by that of the primary galaxies
within the same luminosity range. This calculation assumes that the number of companions per galaxy traces the
number density of galaxies as measured by the luminosity function, and assumes that there is no luminosity-dependent clustering.

For mixed mergers, the above equation is modified into
\begin{eqnarray}\label{G_mix}
1 + G & = &
[\int_{M_{B}^{min}(z)}^{M_{B}^{max}(z)}n_{1}(M,z)dM\int_{M-1.5}^{M+1.5}n_{2}(M',z)dM' \nonumber \\
& & +
\int_{M_{B}^{min}(z)}^{M_{B}^{max}(z)}n_{2}(M,z)dM\int_{M-1.5}^{M+1.5}n_{1}(M',z)dM'] \nonumber \\
& / & [\int_{M_{B}^{min}(z)}^{M_{B}^{max}(z)}n_{1}(M,z)dM\int_{M_{B}^{min}(z)}^{M_{B}^{max}(z)}n_{2}(M',z)dM'
\nonumber \\
& & + \int_{M_{B}^{min}(z)}^{M_{B}^{max}(z)}n_{2}(M,z)dM\int_{M_{B}^{min}(z)}^{M_{B}^{max}(z)}n_{1}(M',z)dM'],
\end{eqnarray}
where $n_{1}$ and $n_{2}$ denote the galaxy number density for blue and red galaxies respectively. We use Equation
\ref{G} and Equation \ref{G_mix} to compute $G$ by adopting galaxy luminosity functions of blue and red galaxies in
the literature \citep[see Table 5 of ][]{fab07}. The value of $G$ is found to range from 0.4 to 1.3, depending on
the galaxy type and the redshift range.

Fig.~\ref{figNm} displays the major merger rates as function of redshift for three types of mergers (wet
mergers,
dry mergers, and mixed mergers). When considering all types of mergers together, it shows that the absolute
merger rate remains fairly constant at $1 \times 10^{-3}$ $h^{3}$Mpc$^{-3}$Gyr$^{-1}$ for $0.1 < z < 1.2$ while the
average wet merger rate is about $7 \times 10^{-4}$ $h^{3}$Mpc$^{-3}$Gyr$^{-1}$ over the same redshift
range.
On the other hand, dry mergers and mixed mergers are found to increase over time. The increase rate of dry
merger
rates is faster than that of the red pair fraction due to the increase in comoving number density of red galaxies
towards lower redshift. It is worth noting that the uncertainty of the absolute merger rates quoted above is
at least a factor of 2 due to the uncertainty in $T_{mg}$.

Fig.~\ref{figNmratio} shows the relative fraction of different mergers in a given redshift bin. At $z >
0.2$, wet mergers
dominate the merger events while dry mergers contribute by a much lesser degree. However, at z = 0.1, the relative
proportions are more similar. The ratio between wet mergers, dry mergers, and mixed mergers is 9:1:3 at $z \sim 1.1$
and 6:5:9 at $z \sim 0.1$, indicating that the role of dry and mixed mergers becomes increasingly important towards
lower redshifts. We also compare our results with the theoretical predictions of the relative fraction of mergers
for different morphological types \citep{kho03}. These predictions are based on semi-analytical galaxy formation
models \citep{kau99,spr01} and merger tree techniques described in \citet{som99}. More details of the models used by
\citet{kho03} can be found in \citet{kho06}. The three colored lines in Fig.~\ref{figNmratio} represent the
model
predictions of the fraction of mergers between early-type galaxies (E-E), late-type galaxies (Sp-Sp), and mixed
mergers (E-Sp) for a field-like environment, corresponding to a dark matter halo of mass $M_{0} \sim 10^{13}
M_{\odot}$. Our observational result is in good agreement with the model predictions by \citet{kho03}, both showing
an increasing fraction of dry and mixed mergers with decreasing redshift since $z \sim 1$.

\section{DISCUSSION}
From the analysis of spectroscopic close pairs, we find that the pair fraction and its evolution depend on the
colors of galaxies. By parameterizing the pair fraction \Nc $\propto~(1+z)^m$, $m$ is found to be $0.41\pm0.2$ for
the full sample of $0.4L^{*} < L < 2.5L^{*}$ galaxies regardless of the companion's color, consistent with the
previous result by \citet{lin04}. It is also in good agreement with the theoretical predictions using pairs of
subhalos \citep{ber06}. Blue galaxies have slightly faster evolution in the blue companion rate (\Ncb) as $m \sim
1.3$ while red galaxies possess inverse evolution with $m \sim -0.9$ in the red companion rate (\Ncr).

Our analysis of \Nc, \Ncb, \Ncr, or \Ncm rules out rapid redshift evolution for $m > 3$ at a 4-sigma level.
This evolutionary trend in the pair fraction can be explained within the context of the observed evolution of the
two-point correlation function \citep{coi08} and the galaxy number density \citep{fab07}. After converting the pair
fraction into galaxy merger rates with the assumed merging fraction in pairs and the merger timescale, we find that
the absolute merger rate is about $1 \times 10^{-3}$ $h^{3}$Mpc$^{-3}$Gyr$^{-1}$ (with a factor of 2
uncertainty) for $0.1 < z < 1.2$. Adopting $h =
0.7$, our estimate is in good agreement with the merger rate $2 - 4 \times 10^{-4}$ Gyr$^{-1}$ Mpc$^{-3}$ obtained
by \citet{lot08} based on morphological approaches. At $z \sim 1.1$, 68\% of mergers are wet, 8\% of mergers are
dry, and 24\% of mergers are mixed, compared to 31\% wet mergers, 25\% dry mergers, and 44\% mixed mergers at $z
\sim 0.1$. Wet mergers dominate merging events at $z = 0.2 - 1.2$, but the relative importance of dry and mixed
mergers increases over time. The good agreement between our observed fraction of various types of mergers and the
predicted results in \citet{kho03} using semi-analytical models supports the importance of the merging hypothesis within
the framework of hierarchical structure formation. In the following sections we discuss several implications of our
results.

\subsection{Mild Evolution or Fast Evolution ?}
The mild evolution of pair fractions and merger rates from  $z \sim 1$ to $0$ is consistent with several
previous studies from either pair counts or morphologies \citep{car00,lin04,lot08}, but disagrees with other recent
works \citep{le00,con03,cas05,kam07,kar07} which claim much higher evolution rates. However, as discussed in
\citet{pat02} and \citet{lin04}, the pair fraction or merger fraction is a function of galaxy luminosity, hence its
evolution depends on how the samples are defined. Moreover, photometric pairs suffer from the contamination by
interlopers, although the spectroscopic pairs may be biased since no spectroscopic survey is complete. Therefore
applying careful projection and completeness corrections in a consistent way across the entire redshift range is
crucial to pin down the true pair/merger fraction. \citet{lot08} also point out that part of the evidence for rapid
evolution in the literature comes from the adopted low pair or merger fraction at $z \sim 0$. In this work, we
select galaxies within a luminosity range such that they evolve in the same way as the $L^{*}$ galaxies out to $z =
1.2$, and apply spectroscopic corrections based on the characteristics of each sample to account for the various
spectroscopic selection effects.

The low-redshift pair fraction at $z \sim 0.1$ obtained here using the MGC sample with 10 \kpc $\leq$ \dis $\leq$
$30$ \kpc and \vel $\leq$ 500 \kms is 5.4\%, which is close to the value of 4.1\% determined independently by
\citet{dep07} using same data set but with slightly different pair selection criteria (\dis $\leq$ $20$ \kpc ; \vel
$\leq$ 500 \kms; -21 $\leq$ \abm - 5log$h$ $\leq$ -18). Both of these results are higher than the pair fraction of
the SDSS sample reported by \citet{kar07} and \citep{bel06b}. Future works calibrating the merger fraction at low
redshifts \citep{pat08} will help to disentangle the issue of different evolutionary trends \citep[also see the
discussion in][]{lot08}.

\subsection{The Accumulated Merger Fraction since $z \sim 1.2$}
To determine the accumulated effect of major mergers on galaxies at the present epoch, we calculate the fraction of
present day galaxies that have undergone major mergers since $z \sim 1.2$.  We consider two cases here: one has
mergers among galaxies within $-21 <$ \mbe $< -19$ (i.e., the luminosity of both pair components is within $0.4L^{*}
< L < 2.5L^{*}$); the other is for galaxies with $-21 <$ \mbe $< -19$ that merge with companions with luminosity
ratios ranging between 4:1 and 1:4. In the first case, we follow Equation (32) in \citet{pat00}:

\begin{equation}\label{frem}
f_{rem} = 1 - \prod_{j=1}^N{1- C_{mg}N_{c}(z_j) \over 1 - 0.5 C_{mg}N_{c}(z_j)},
\end{equation}
where $f_{rem}$ is the remnant fraction, $C_{mg}N_{c}$ gives the fraction of galaxies that will undergo mergers during the
time interval $T_{mg}$, and $z_{j}$ corresponds to a look-back time of t = $jT_{mg}$.
%The resulting remnant fraction depends again on the definition of major mergers.
Adopting $C_{mg} = 0.6$ as used in \S 3.2, and $N_{c}(z)$ from the fit of the data for
the case of $r_{max}$ = 30 \kpc, our result implies that 22\% of today's galaxies with $0.4L^{*} < L < 2.5L^{*}$  have
experienced mergers with galaxies within the same luminosity range since $z \sim 1.2$.

In the second case which is a better probe of the major merger rate, Equation~\ref{frem} needs to be modified as:
\begin{equation}\label{frem2}
f_{rem} = 1 - \prod_{j=1}^N{1- (1+G)C_{mg}N_{c}(z_j) \over 1 - 0.5 C_{mg}N_{c}(z_j)},
\end{equation}
where the term $(1+G)$ accounts for the missed companions (see \S 3.2). Since the factor $G$ is approximately 1
at all redshifts, we conclude that about 54\% of $0.4L^{*} < L < 2.5L^{*}$ galaxies have undergone a major merger
since $z \sim 1.2$, but this depends sensitively on the assumed merger timescale. Both estimates above of the
remnant fraction are about 2-6 times greater than that reported in \citet{lin04}. The major cause of this
difference can be traced to the different definitions of major mergers as well as the different choice of pair
separations when calculating the remnant fraction. \footnote{\citet{lin04} calculated the remnant fraction by using
pairs with $r_{max}$ = 20 \kpc and by requiring both merger components to have $-21 <$ \mbe $< -19$.}

\subsection{The Roles of Dry Mergers in the Formation of Massive Galaxies}

Previous works have suggested that dry mergers are likely responsible for the growth of massive galaxies on the red
sequence \citep{bel04,fab07}, because of the lack of blue galaxies massive enough to migrate from the blue cloud to
the red sequence. Moreover, there is observational evidence showing the nonnegligible amount of dry mergers
occurring in the past 8 Gyr \citep[van Dokkum 2005; Bell et al. 2006a, White et al. 2007, although see][]{mas06,sca07}, as well as evidence from
theoretical expectations \citep{kho03,naa06,cat08} . We now discuss the implication of our merger studies on this topic. We have shown that at a given
luminosity range, the merger events are always dominated by wet mergers, followed by mixed mergers, and then dry
mergers in terms of the event rates over the redshift range $0.2 < z < 1.2$. This is mainly because the number
density of blue galaxies dominates the total galaxy population in the luminosity range we consider. However, given
the fact that the pair fraction and merger rate also depend on the clustering properties of galaxies in addition to
the galaxy number density, the probability of red galaxies having a red companion \Ncr turns out to be comparable to
the blue companion rate for blue galaxies \Ncb at $z<1.2$. At $z < 0.4$, \Ncr becomes even greater than \Ncb as
shown in Fig.~\ref{figNc} and Table~\ref{tab1}. That is to say, the probability of mergers within the red
sequence is
greater than that within the blue cloud at low redshifts.

We can compare our results to previous attempts at measuring the dry merger frequency. Our derived \Ncr
average over $0.1 < z < 0.7$ is similar to the fraction of dry-merger candidates $\sim$ 3\% found by \citet{bel06a}
from the GEMS survey. In addition, our fitted dry pair fraction at $z \sim 0$ is about 0.045 (see
Table~\ref{tab2}), which is also in broad agreement with the companion fraction ($\sim 0.06$)
within the red sequence estimated by \citet{van05} using nearby galaxy samples. On the other hand,
\citet{mas06}
found very
small merger rates for luminous red galaxies, on the order of $0.6 \times 10^{4}$ Gyr$^{-1}$Gpc$^{-3}$ from the SDSS
Luminous Red Galaxy sample (LRG). Their finding is about 23 times lower than our estimates of dry merger rates at $z
\sim 0.1$ and 7 times lower than ours at $z \sim 0.3$ \footnote{Note that their merger rates are quoted using the
unit Gpc$^{-3}$Gyr$^{-1}$ while our results in Table~\ref{tab1} are in $h^{3}$Mpc$^{-1}$Gyr$^{-1}$. We adopt
$h = 0.7$ when
doing comparisons.}. The major discrepancy can be attributed to the different luminosity ranges being sampled: their
choice of faint-end magnitude is brighter than typical $L^{*}$ galaxies by almost 2 mag while ours is fainter than $L^{*}$ by 1 mag.
Since the luminosity function of red galaxies shows a strong decline towards the bright end, we expect that mergers occurring among luminous red galaxies should be
much lower than that among less-luminous ones. This effect is also found by \citet{pat02} and \citet{lin04}.

In order to assess how much dry mergers with luminosity ratios less than 4:1 may contribute to the growth of massive
red galaxies, we compute the following quantities, $<$ \Nmgd$>$$T_{z}$/$n_{g}^{r}(0)$, where $<$ \Nmgd$>$ is the
average dry merger rate over $0 < z < 1$, $T_{z}$ is the cosmic time since $z = 1$, and $n_{g}^{r}$ is the number
density of red galaxies within a 2 mag bin centered on $L^{*}$ at $z = 0$. Taking $<$ \Nmgd$>$ as
 2 $\times$ $10^{-4}$ $h^{3}$Mpc$^{-1}$Gyr$^{-1}$, $T_{z}$ as 8 Gyr, and $n_{g}^{r}(0) \sim 6.7 \times 10^{-3}
$ $h^{3}$Mpc$^{-1}$, we find that about 24\% of present red galaxies have experienced dry mergers. Our result is
slightly lower than the 35\% found by \citet{van05}, mainly because they have assumed a constant dry merger
rate over time
while we find a decreasing rate of dry mergers with increasing redshift.

\subsection{The Role of Mixed and Wet Mergers vs. Dry Mergers}

The contribution of mixed mergers and wet mergers to the formation of red galaxies is less straightforward to
constrain given the difficulty in handling how often and how soon the mixed and wet mergers transform the merger
remnant into a red sequence galaxy. Under the extreme assumption that both mixed mergers and wet mergers lead to the
formation of red galaxies immediately, the fraction of present day red galaxies that have experienced mixed and wet
mergers is roughly 36\% and 71\% respectively. These values are certainly over-estimated since not necessarily all
remnants of mixed or wet mergers end up in the red sequence. We note that at a fixed luminosity, the stellar mass of
red galaxies is systematically higher than that of blue galaxies (see Equation 1 of Lin et al. 2007); in other
words, the stellar mass of the progenitors of dry mergers is larger than that of wet mergers or mixed mergers in our
sample since we select the pairs based on the luminosity cut in both the blue cloud and red sequence. For example,
the typical stellar mass of our selected blue galaxies is $\sim 2\times10^{10}M_{\odot}$ and that of our red
galaxies is $\sim 10^{11}M_{\odot}$. Hence, the merger remnant of these three types of mergers being considered in
our sample will likely end up as red galaxies in different stellar mass regimes. A plausible scenario is that the star
formation is quenched after the process of wet mergers and/or mixed mergers, resulting in the formation of some
portion of the red galaxies with intermediate masses. The massive red galaxies are then built up through dry mergers
between galaxies with intermediate masses at a later time since our results suggest that dry-mergers play an
increasing role at lower redshifts.

\section{CONCLUSION}
Combining the DEEP2, TKRS, MGC, CNOC2, and SSRS2 catalogs, we study
the redshift evolution of the pair fraction and major merger rates of
wet, dry, and mixed mergers for galaxies with $-21 <$ \mbe $< -19$ out
to $z \sim 1.2$. The merger candidates are identified as close pairs
based on their projected separation on the sky and relative
line-of-sight velocities. Wet, dry, and mixed mergers are classified according to the colors of the individual components in close pairs. Our results can be summarized as follows:

1. Parameterizing the evolution of the pair fraction as $(1+z)^{m}$, we find that $m = 0.41\pm0.20$ for the full sample, consistent with the low value of $m$ as found by \citet{lin04}.

2. The values of $m$ depend on the color combination in close pairs. Blue
galaxies show slightly faster evolution in the blue companion rate with $m = 1.27\pm0.35$ while red galaxies have
had fewer red companions in the past as evidenced by the negative slope $m = -0.92\pm0.59$. On the other hand, $m = -1.52\pm0.42$ for mixed pairs. The different trends of
the pair fraction evolution are consistent with the predictions from the observed evolution of galaxy number densities
and the two-point correlation function for both the blue cloud and red sequence.

3. For the chosen luminosity range, we find that at low redshift ($z < 0.4$) the pair fraction within the red sequence is greater than that of the blue cloud, indicating a higher merger probability within the red sequence compared to that within the blue cloud.

4. With further assumptions on the
merger timescale and the fraction of pairs that will merge, the galaxy major merger rates for $0.1 < z <
1.2$ are estimated to be
$\sim~10^{-3}$ $h^{3}$Mpc$^{-3}$Gyr$^{-1}$ (with the uncertainty about a factor of 2),
dominated by wet mergers (gas-rich mergers) until
very recently. There were more wet merger events than dry or mixed mergers because of the higher number
density of blue galaxies for the
chosen luminosity cut. However, the fraction of mergers which are dry or mixed increases over time, from 8\%
and 24\% at $z \sim 1$ to 25\% and 44\% at
$z = 0$ respectively.
The growth of dry merger rates with decreasing redshift is mainly due
  to the rise in the co-moving number density of red galaxies with time. Our results on the
fraction of mergers of different types are in good agreement with theoretical predictions by \citet{kho03} based on
semi-analytical models.

5. About 22\% to 54\% of present-day $L^{*}$ galaxies have
experienced major mergers since $z \sim 1.2$, depending on the definition of major mergers. Moreover, 24\% of the
red galaxies at the present epoch have had dry mergers with luminosity ratio less than 4:1 since $z \sim 1$.

6. Given the $B$-band luminosity cut, the blue and red galaxies in our sample possess different stellar masses: the typical stellar mass of red galaxies in our sample is about 5 times greater than that of the blue galaxies. By assuming that a significant fraction of
wet/mixed mergers will end up as red galaxies as well as dry mergers, our results suggest that the three types of mergers lead to red galaxies in
different stellar mass regimes: the wet mergers and/or mixed mergers may be partially responsible for producing red galaxies
with intermediate masses while dry mergers in our
  sample produce a significant portion of massive red galaxies at low
  redshift.\\

We have demonstrated that the redshift evolution of pair fractions and merger rates depend on galaxy types, owing to the color dependence of the evolution in the galaxy number density and clustering. In terms of the absolute number of events, wet mergers dominate merger events at $0.2 < z < 1.2$. However, dry and mixed mergers become more important over time, in particular at very low redshifts ($z < 0.2$). Our findings support the late growth of massive red galaxies through dry mergers
as concluded by \citet{van05} and \citet{bel06a}. Moreover, our results also suggest that wet and mixed mergers are responsible for producing red-sequence galaxies in lower
stellar mass regimes. More observational and theoretical studies on the effects of the three
types of mergers on the star formation in the merger remnants will help us to constrain better the role of galaxy
mergers in forming red-sequence galaxies. One uncertainty that can affect the relative importance of the three types of mergers is that we have adopted a constant merger timescale for all types of mergers in our analysis. This approach is based on the assumption that the $B$-band light is a good tracer of dynamical mass \citep{kan08} and hence the merger timescale should be comparable in different types of pairs selected by the same $M_{B}$ cut. Future works to pin down the merger timescale more precisely will be valuable in determining the relative importance among wet, dry, and mixed mergers.

\acknowledgments We thank the referee for
a very thorough and helpful report. This work was supported by NSF grants AST00-71198, AST05-07428 and AST05-07483, an NSERC Discovery
Grant to D. R. P, and an NSC grant NSC95-2112-M-002-013 to T. Chiueh. We thank Olivier Le F\`evre, Alison Coil,
Arjun Dey, Youjun Lu, and Sheila Kannappan for useful discussions; S. Khochfar and A. Burkert for providing data from their previous theoretical works and for helpful comments; the Taiwan CosPA project for accessing CFHT Observing time; the staff of CFHT
for conducting MegaCam Observations and image pre-processing; Laurent Domisse of the Terapix team for data reduction and
stacking; and David Gilbank and Alex Conley for providing defringing program to process the MegaCam images. This work is
based in part on data products produced at the TERAPIX data center located at the Institut d'Astrophysique de Paris.

The DEEP2 Redshift Survey has been made possible through the dedicated efforts of the DEIMOS instrument team  at UC
Santa Cruz and support of the staff at Keck Observatory.

The Millennium Galaxy Catalogue consists of imaging data from the Isaac Newton Telescope and spectroscopic data from
the Anglo Australian Telescope, the ANU 2.3m, the ESO New Technology Telescope, the Telescopio Nazionale Galileo and
the Gemini North Telescope. The survey has been supported through grants from the Particle Physics and Astronomy
Research Council (UK) and the Australian Research Council (AUS). The MGC data and data products are publicly
available from http://www.eso.org/~jliske/mgc/ or on request from J. Liske or S.P. Driver.

We close with thanks to the Hawaiian people for use of their sacred mountain.

\begin{deluxetable}{crrrrrrrrr}
 \tabletypesize{\scriptsize}
 \tablecaption{Pair Statistics Using $r_{max}$ = 30 \kpc and Comoving Volume Merger Rates \label{tbl-1}}
 \tablewidth{0pt}
 \tablehead{ \colhead{Sample} & \colhead{$\overline{z}$} & \colhead{\Nc} & \colhead{\Ncb} &\colhead{\Ncr} &\colhead{\Ncm}& \colhead{$N_{mg}$} & \colhead{$N_{mg}^{wet}$} & \colhead{$N_{mg}^{dry}$} & \colhead{$N_{mg}^{mix}$} }
 \startdata

 DEEP2 Field 1  &0.623 &0.047$\pm$0.011 &0.027$\pm$0.009 &0.029$\pm$0.012 &0.019$\pm$0.007 &1.02E-03 &4.55E-04 &1.20E-04 &3.35E-04\\
(EGS) &0.875 &0.048$\pm$0.010 &0.041$\pm$0.009 &0.025$\pm$0.014 &0.010$\pm$0.004 &1.10E-03 &7.88E-04 &7.39E-05 &1.88E-04\\
 \nodata &1.094 &0.061$\pm$0.015 &0.050$\pm$0.013 &0.076$\pm$0.052 &0.008$\pm$0.006 &1.13E-03 &7.31E-04 &2.21E-04 &1.09E-04\\
\tableline
 DEEP2 Field 3  &0.883 &0.041$\pm$0.008 &0.030$\pm$0.007  &0.031$\pm$0.015  &0.011$\pm$0.005 &5.71E-04 &3.42E-04 &5.91E-05 &1.22E-04\\
 \nodata &1.084 &0.088$\pm$0.018 &0.062$\pm$0.014 &0.053$\pm$0.037 &0.027$\pm$0.013 &1.54E-03 &8.22E-04 &1.63E-04 &3.64E-04\\
\tableline
 DEEP2 Field 4  &0.874 &0.044$\pm$0.009 &0.040$\pm$0.010 &0.026$\pm$0.011 &0.007$\pm$0.003 &8.03E-04 &5.93E-04 &7.04E-05 &1.04E-04\\
 \nodata &1.083 &0.083$\pm$0.023 &0.081$\pm$0.025 &0.000$\pm$0.000 &0.013$\pm$0.011 &1.25E-03 &8.88E-04 &0.000E+00 &1.53E-04\\
\tableline
 TKRS &0.575 &0.063$\pm$0.014 &0.034$\pm$0.014 &0.019$\pm$0.009  &0.034$\pm$0.010 &1.45E-03 &5.94E-04 &8.64E-05 &6.29E-04\\
 \nodata &0.874 &0.060$\pm$0.014 &0.039$\pm$0.013 &0.074$\pm$0.038 &0.014$\pm$0.007 &1.62E-03 &9.00E-04 &2.54E-04 &2.86E-04\\
 \nodata &1.056 &0.159$\pm$0.050 &0.092$\pm$0.037 &0.000$\pm$0.000 &0.076$\pm$0.039 &2.21E-03 &1.06E-03 &0.000E+00 &8.11E-04\\
\tableline
 SSRS2 &0.014 &0.028$\pm$0.006 &\nodata &\nodata &\nodata &\nodata &\nodata &\nodata &\nodata\\
\tableline
 MGC &0.120 &0.054$\pm$0.005 &0.023$\pm$0.004 &0.043$\pm$0.006 &0.024$\pm$0.003 &1.60E-03 &4.98E-04 &4.00E-04 &7.20E-04\\
\tableline
 CNOC2 &0.335 &0.041$\pm$0.011   &0.025$\pm$0.011  &0.032$\pm$0.016 &0.013$\pm$0.007 &9.55E-04 &5.41E-04 &1.37E-04
 &2.47E-04

 \enddata
 \tablecomments{Here \Nc is the companion rate per galaxy; \Ncb is the blue companion rate per blue galaxy; \Ncr denotes
the red companion rate per red galaxy; \Ncm gives the average number of companions with colors opposite that of the
primary galaxy. \Nmg, \Nmgw, \Nmgd, \Nmgm are the comoving merger rate for total mergers, wet mergers, dry mergers,
and mixed mergers in units of number of mergers $h^{3}$ Mpc$^{-3}$Gyr$^{-1}$. \label{tab1}}
\end{deluxetable}

\begin{deluxetable}{crrrrrr}
 \tabletypesize{\scriptsize}
 \tablecaption{Results of Fitting Parameters for The Pair Fraction \label{tab2}}
 \tablewidth{0pt}
 \tablehead{ \colhead{Pair Types} & \colhead{$N_{c}(0)^{30}$} & \colhead{$m^{30}$} & \colhead{$N_{c}(0)^{50}$} &\colhead{$m^{50}$} &\colhead{$N_{c}(0)^{100}$}& \colhead{$m^{100}$} }
 \startdata

 All &0.041$\pm$0.004 &0.41$\pm$0.20 &0.084$\pm$0.006
 &0.41$\pm$0.14 &0.210$\pm$0.004 &0.29$\pm$0.05\\
\tableline
 Blue-Blue &0.018$\pm$0.004 &1.27$\pm$0.35 &0.029$\pm$0.004
 &1.32$\pm$0.28 &0.082$\pm$0.007 &1.09$\pm$0.15\\
\tableline
 Red-Red &0.045$\pm$0.009 &-0.92$\pm$0.59 &0.099$\pm$0.013
 &-0.51$\pm$0.40 &0.262$\pm$0.021 &-0.85$\pm$0.23\\
\tableline
 Mixed &0.029$\pm$0.005 &-1.52$\pm$0.42 &0.048$\pm$0.006
 &-0.64$\pm$0.26 &0.099$\pm$0.008 &-0.29$\pm$0.16

 \enddata
 \tablecomments{Here $N_{c}(0)$ and $m$ are the fitting parameters for the evolution of pair fraction in the form of $N_{c}
 = N_{c}(0)(1+z)^{m}$. The superscripts 30, 50 and 100 denote the values of $r_{max}$ in unit of \kpc used to
select close pairs.}
\end{deluxetable}

\end{document}